\documentclass[10pt,conference]{IEEEtran}
\usepackage{cite}
\usepackage{amsmath,amssymb,amsfonts}
\usepackage{algorithmic}
\usepackage{graphicx}
\usepackage{textcomp}
\def\BibTeX{{\rm B\kern-.05em{\sc i\kern-.025em b}\kern-.08em
    T\kern-.1667em\lower.7ex\hbox{E}\kern-.125emX}}
\usepackage{paralist}

\usepackage{xspace}
\makeatletter
\DeclareRobustCommand\onedot{\futurelet\@let@token\@onedot}
\def\@onedot{\ifx\@let@token.\else.\null\fi\xspace}
\def\eg{\emph{e.g}\onedot}
\def\ie{\emph{i.e}\onedot}
\def\etc{\emph{etc}\onedot}

\usepackage{csquotes} %
\usepackage[colorlinks=true,linkcolor=black,anchorcolor=black,citecolor=black,filecolor=black,menucolor=black,runcolor=black,urlcolor=black]{hyperref} %

\begin{document}

\title{Software Resurrection: Discovering Programming Pearls by Showing Modernity to Historical Software}

\author{
  \IEEEauthorblockN{Abhishek Dutta}
  \IEEEauthorblockA{\textit{\url{https://abhishekdutta.org/sr/}}}
}

\maketitle

\begin{abstract}
Reading computer program code and documentation written by others is,
we are told, one of the best ways to learn the art of writing
intelligible and maintainable code and documentation. The software
resurrection exercise, introduced in this paper, requires a motivated
learner to compile and test a historical release (e.g. 20 years old)
version of a well maintained and widely adopted open source software
on a modern hardware and software platform. This exercise concludes by
writing a critique based on issues encountered while compiling and
testing a historical software release on a hardware and software
platform that could not have been foreseen at the time of release. The
learner is also required to fix the issues as a part of the software
resurrection exercise. The seemingly pointless exercise of
resurrecting a historical software allows motivated learners to
experience the pain and joy of software maintenance which is essential
for understanding the factors that contribute to intelligibility and
maintainability of program code and documentation. The concept of
software resurrection exercise is illustrated using a version of the
SQLite database engine that was released 20 years ago. This
illustration shows that software engineering principles (or
programming pearls) emerge when a historical software release is
adapted to run successfully on a modern platform. The software
resurrection exercise also has the potential to lay foundations for a
lifelong willingness to explore and learn from existing program code.
\end{abstract}

\begin{IEEEkeywords}
software resurrection, programming pearls, programming wisdom,
intelligible code and documentation, software maintenance
\end{IEEEkeywords}

\section{Introduction}
This paper introduces the concept of \textit{software resurrection} as
an exercise for discovering software engineering principles that helps
create intelligible and maintainable program code and
documentation. The exercise is pursued by a motivated learner who is
already familiar with computer programming and wishes to learn the art
of writing program code and documentation that is easy to understand
and requires less maintenance. The exercise is carried out on a
historical release (\eg released 20 years ago) of a well maintained
and widely adopted software. The selected historical software release
should include a self contained suite of tests as well as be written
in a programming language that is familiar to the learner.

\begin{figure}[ht]
  \centering
  \includegraphics[width=\linewidth]{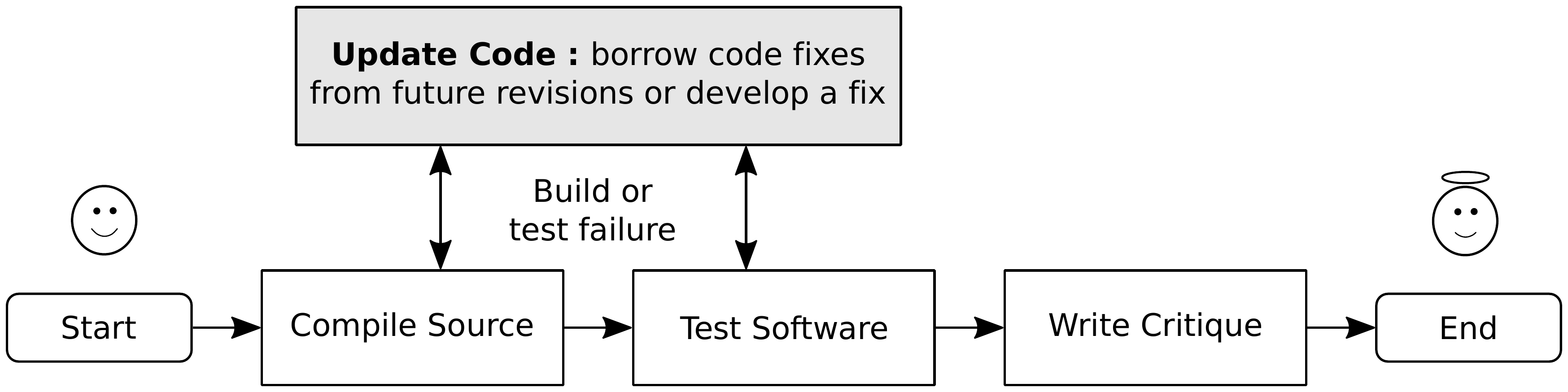}
  \caption{Software resurrection exercise begins with compilation of
    an old release of a well maintained and widely used software in a
    modern hardware and software platform. After successful
    compilation, the software's functionality is verified using
    automated suite of tests included with the release. Learning
    opportunities are provided by failure in compilation and testing
    processes. Learners engage with the program code and documentation
    to develop a fix for these issues. Finally, the exercise concludes
    by writing a critique of the software code and documentation which
    provides the opportunity to reflect on the experiences of compiling
    and testing the software in a modern platform.}
  \label{fig:software-resurrection-flowchart}
\end{figure}

The software resurrection exercise consists of three stages as shown
in \figurename~\ref{fig:software-resurrection-flowchart}. The exercise
begins with the compile stage which requires the learner to compile
the historical software release on a modern platform. The modern
platform, for example, can be a 64 bit multiple core x86 machine
running a latest version of Debian Linux with a latest version of a
compiler like GCC or Clang. The compilation process may requiring
building the dependency libraries. After the compilation succeeds, the
next stage requires the learner to test the compiled software by
executing the self contained tests on the same modern platform. The
software resurrection exercise concludes with a critique written by
the learner which is similar in spirit to the practice of literary
criticism in English literature. The critique contains a brief
description of the issues encountered during the compilation and the
test stages and a detailed description of the fix developed by the
learner. The critique allows the learner to reflect and explore if the
issues identified during the software resurrection exercise points
towards some underlying software engineering principles.

How can, one may wonder, the seemingly pointless activity of
compiling, testing and critiquing an old software on a modern platform
lead to discovery of software engineering principles? The compile and
test activities are likely to fail because the developers of the
historical software release could not have foreseen the features and
constraints of a modern hardware and software platform. These failures
reveals some important facet of software engineering and provides a
thread of investigation to the learner. For example, if the modern
compilers have dropped support for a non-standard feature that was
widely used and supported 20 years ago then a historical software
relying on such non-standard feature would not compile on a modern
platform thereby revealing the cost of relying on non-standard
features of a compiler. The software resurrection exercise also
requires the learner to develop a fix for any issues encountered
during the exercise. To fix an issue, the learner must read and
understand the program code and documentation contained in the
historical software release. This provides the learner with a first
hand experience of software maintenance. For example, if the program
code is well structured and clearly documented, it will simplify the
process of fixing the issue caused by compilers dropping support for a
non-standard feature. On the other hand, a convoluted code structure,
missing documentation, unintelligible identifier names, \etc will
puzzle the learner and demand more time and require significant effort
to develop a fix. Such joys and frustrations are essential to learn
the aspects of program code and documentation that survives the test
of time and remains intelligible even after many years. Such
experiences encourages a motivated learner to adopt best practices in
software engineering that brings joy to a maintainer and avoid the
aspects of program code and documentation that are difficult to
understand or maintain.

The software resurrection exercise also requires the learner to write
a critique of the software which provides an opportunity to reflect on
the experiences gathered during the compilation and testing
stages. Every issue encountered by the learner points to an assumption
or a decision made by the developers of the historical software
release several years ago. The critique activity encourages the
learner to not only identify those assumptions and decisions but also
develop an understanding of the circumstances under which those
assumptions and decisions were made. For example, the decision to rely
on a non-standard feature of a compiler was, most likely, based on the
assumption that such features will always be supported by the
compilers. However, such assumptions may not hold after 20 years and
therefore would result in compilation or testing failure. To
understand the circumstances under which this decision was made, the
learner has to investigate about the alternative options that were
available at that time and if choosing one of the alternative options
would have avoided the failure. Furthermore, the learner could also
investigate if it would have been wiser to not have a software feature
that is based on a non-standard feature of the compilers of that
age. Such reasoning and reflections allows the learner to truly
understand the factors that contributed to the failure. This helps
develop an impartial view towards a software engineering practice.

There is no quantitative experimental data yet to support the claim
that the proposed software resurrection exercise allows a learner to
discover software engineering principles behind intelligible and
maintainable program code and documentation. Readers are encouraged to
pursue the software resurrection exercise and self evaluate their
learning experience. Philosophy provides some insights into the
effectiveness of the software resurrection exercise which allows
modern day developers to break free ``from the tyranny of the here and
the now''~\cite[p.162]{russel1956portraits} by introducing them to
program code and documentation that are remote in time. Such forays
into historical software releases enables a learner to view things
(\eg software engineering practices) from different perspectives. The
learner is no longer captive of their personal viewpoints and becomes
capable of surveying a wider horizon of ideas thereby contributing to
growth in their wisdom. It is this wisdom that emerges as programming
pearls that are commonly shared by experienced programmers who get to
know about many things that are remote in time or space by the virtue
of their long careers spanning various application domains. Bertrand
Russel, a 20th century philosopher, remarked that wisdom can be
learned by such excursions into ``things that are somewhat remote in
time or space''.

Section~\ref{sec:sr-of-sqlite} shows an example of the software
resurrection exercise pursued on a version of the SQLite database
engine that was released 20 years ago. The compile and test stages are
described in Section~\ref{sec:compile} and \ref{sec:test}
respectively. The critique stage is described in
Section~\ref{sec:critique}. Additional examples of software
resurrection exercise are available
online\footnote{\url{https://abhishekdutta.org/sr/}}.
Section~\ref{sec:related_work} discusses other work that are related
to the concept of software resurrection. The conclusions from this
research is presented in Section~\ref{sec:conclusion}.

\section{Software Resurrection of SQLite-2002}
\label{sec:sr-of-sqlite}
\href{https://www.sqlite.org/}{SQLite} is a light weight and portable
database engine that has been actively developed since the year 2000
and has seen wide adoption by users. Its program source code is
dedicated to the public domain which entails complete freedom to use
the program code for any purpose; this paper uses it for learning. The
version \href{https://www.sqlite.org/src/info/61c38f3bfef430f3}{2.2.1}
of SQLite released in the year 2002 -- henceforth referred as
\textit{sqlite-2002} -- has been selected as the historical release for the
software resurrection exercise because it
\begin{itemize}
\item is sufficiently remote in time (\i.e. 20 years old),
\item has a publicly accessible version controlled history of all
  code revisions, and
\item includes self contained suite of tests to verify its functionality.
\end{itemize}

The sqlite-2002 is compiled and tested on the following hardware and
software platform as described in Section~\ref{sec:compile} and
Section~\ref{sec:test} respectively.
\begin{itemize}
\item \textit{Hardware} : Dell XPS 15 laptop purchased in 2019
  containing Intel i9-9980HK CPU @ 2.40GHz (x86\_64, Little Endian)
  with address sizes of 39 bits physical(48 bits virtual) and 16 CPUs.
\item \textit{Software} : Debian GNU/Linux
  \href{https://www.debian.org/News/2022/20220709}{11.4 (bullseye)}
  operating system released on 9th July 2022 with Linux Kernel
  5.10.0-16-amd64 and a build system comprising of gcc-10.2.1, GNU Make
  4.3 and GNU Autoconf 2.69.
\end{itemize}

The software resurrection of sqlite-2002 concludes with a critique, an
example of which is presented in Section~\ref{sec:critique}. Some of
the details have been omitted from the description of compilation and
testing stages in order to improve the readability of this paper; full
details are included in the online version.

\subsection{Compile Source}
\label{sec:compile}
The sqlite-2002 (i.e. sqlite-2.2.1 release) is downloaded and compiled
using the standard autoconf based \texttt{./configure} and
\texttt{make} commands. The first build issue is related to a breaking
change introduced by the GCC compiler.

{\small
\begin{verbatim}
varargs.h:4:2: error:
  #error "GCC no longer implements ."
varargs.h:5:2: error:
  #error "Revise your code to use ."
...
sqlite/tool/lemon.c:1096:1: error:
  expected declaration specifiers before
  ‘va_dcl’
\end{verbatim}
}

\subsubsection{Compiler Drops Support}
\label{compile:issue1}
The sqlite-2002 does not compile in gcc-10.2.1 (2021) and autoconf
2.69 (2012) because the SQL statement parser defined in
\texttt{tool/lemon.c} uses \texttt{varargs.h} header file which was
deprecated by the gcc compiler since 4.0 (2005) release. The gcc
compiler dropped support for \texttt{varargs.h} since April 2004 and
switched to supporting \texttt{stdarg.h} header file to provide the
same functionality. The sqlite developers must have adapted their code
before the compilers implemented this breaking change. Therefore,
version control history of sqlite should contain a fix in one of the
future revisions. The vararg issue was fixed only in
\href{https://www.sqlite.org/src/info/590f963b6599e4e2}{sqlite-2.8.1}
release by replacing dependence on \texttt{varargs.h} with
\texttt{stdarg.h}. Unfortunately, a fix for this issue did not appear
in a single version control revision (or commit) and the code updates
have to be selectively borrowed from the
\href{https://www.sqlite.org/src/info/590f963b6599e4e2}{sqlite-2.8.1}
release.

\subsubsection{Name Conflict with Standard Library}
\label{compile:issue2}
After resolving the \texttt{varargs.h} issue, the compilation proceeds
ahead and reveals the second issue caused by naming conflict with the
standard library.

{\small
\begin{verbatim}
../sqlite/src/shell.c:50:14: error:
  conflicting types for 'getline'
  | static char *getline(char *zPrompt, ...){

/usr/include/stdio.h:616:18: note:
  previous declaration of 'getline' was here
  | extern __ssize_t getline (char ** ...)
\end{verbatim}
}

The error message informs that the \texttt{getline()} method has been
declared by the standard library as well as the \texttt{src/shell.c}
sqlite source. If the \texttt{getline()} method were a part of the
standard library at the time of release, the authors would have
renamed their version of \texttt{getline()} before the release to
avoid such conflicts. Therefore, the standard library must have been
updated after the release of sqlite-2002. It is highly likely that one
of the code revisions (or checkout) in the version control history of
SQLite may contain a fix for this issue as the SQLite software would
have adapted to this change in the standard library. A search of the
version control system of sqlite for the keyword ``getline()'' returns
only one result which corresponds to the
\href{https://sqlite.org/src/info/558969ee8697180c}{revision} that
resolved the name conflict. The conflict resolution involved renaming
the method to \texttt{local\_getline()}. This wise decision has
ensured that the renamed method has not required such changes in more
than 20 years since the change. The sqlite-2002 source compiles
successfully after applying the
\href{https://sqlite.org/src/vpatch?from=93a2c961b17d2459&to=558969ee8697180c}{patch}
generated from the sqlite public version control repository.

\subsection{Run Tests}
\label{sec:test}
The sqlite-2002 is tested using the standard autoconf based
\texttt{make test} command. An output like ``All tests passed''
generated by the test command provides assurances that the software
behaves in an expected way. However, the test command fails to compile
because the Tcl library required to build the tests is missing. The
autoconf's configure script -- created in 2002 -- is responsible for
locating all the dependencies required to compile the tests. This
script is unable to recognise the more recent version of Tcl library
that is installed using the operating system's package
manager. Therefore, the script that compiles all the tests (\ie
\texttt{Makefile} which gets generated by the configure script) is
manually updated such that the \texttt{TCL\_FLAGS} and \texttt{LIBTCL}
variables point to the Tcl library installed by the operating system.

\subsubsection{Breaking Changes Introduced by a Dependency}
\label{test:issue1}
The build system is able to locate the Tcl library. However, the
latest version of Tcl library appears to be incompatible as revealed
by the following compilation error.

{\small
\begin{verbatim}
sqlite/src/tclsqlite.c:622:36: error:
  "Tcl_Interp" has no member named "result"
  | if( zInfo==0 ) zInfo = interp->result;
\end{verbatim}
}

The error message indicates that the Tcl library has introduced a
breaking change because of which the \texttt{result} field is not
available in the \texttt{Tcl\_Interp} data structure. The
\texttt{Tcl\_Interp} API documentation
\href{https://www.tcl.tk/man/tcl/TclLib/Interp.html}{describes this
  breaking change} and requires users of this legacy feature to define
the \texttt{USE\_INTERP\_RESULT} macro in order to enable access to
the \texttt{result} field. This issue gets resolved by defining the
required macros as advised by the API documentation. The tests compile
successfully after an \texttt{extern} qualifier is added to the
declaration of a variable flagged as undefined by the compiler.

\subsubsection{A 32 Bit Software in a 64 Bit System}
\label{test:issue2}
Tests compile successfully but the tests fail to execute on a modern
platform due to a \texttt{SEGFAULT} error.

{\small
\begin{verbatim}
./testfixture ../sqlite/test/quick.test
bigrow-1.0... Ok
bigrow-1.1... Ok
...
btree-1.4.1... Ok
btree-1.5...
  make: [Makefile:232: test] Segmentation fault
\end{verbatim}
}

The \texttt{SEGFAULT} errors are caused by programs trying to access a
memory location that it is not allowed to access. The program code
that is causing this error can be located using the GNU Debugger (gdb)
backtrace functionality.

{\small
\begin{verbatim}
$ gdb --args .libs/lt-testfixture ...quick.test
(gdb) run
...
btree-1.4.1... Ok
btree-1.5...
Program received signal SIGSEGV
(gdb) backtrace
#0  sqliteBtreeCursor (pBt=0x555e3db0, ...)
    at ../sqlite/src/btree.c:823
#1  btree_cursor (argv=0x555555588a80, ...)
    at ../sqlite/src/test3.c:527
#2  btree_cursor (argv=0x555555588a80, ...)
    at ../sqlite/src/test3.c:506
...
#8  main (argv=0x7fffffffdff8, ...)
    at ../sqlite/src/tclsqlite.c:620
\end{verbatim}
}

The backtrace output shows that the pointer address for \texttt{argv}
variable is 64 bit long (\ie \texttt{0x555555588a80}) while the
pointer address \texttt{pBt} is only 32 bits long (\ie
\texttt{0x555e3db0}). An arduous debugging session reveals that the
\texttt{SEGFAULT} is caused by the program code that incorrectly
converts the btree pointer address to string representation by wrongly
assuming that memory addresses are 32 bits long. This assumption was
true in the year 2002 when the memory could conveniently be
represented by only 32 bits. In a modern 64 bit platforms, memory
addresses are represented by 64 bits (\ie 8 bytes). This issue
requires fix in two places: first when a pointer address is converted
to string representation and second when the string representation is
converted back to pointer address. The string representations are used
by the Tcl script to operate on a test database. To address the first
issue, the \texttt{\%p} format specifier (instead of \texttt{\%x}
which assumes 32 bit argument) is used to represent the 64 bit pointer
address as string. The second issue is addressed by using
\texttt{strtol()} function to convert back the string representation
to the pointer address as shown below. Such fixes have to be applied
at multiple places in the following source files:
\texttt{src/\{test1.c, test2.c, test3.c\}}.

{\small
\begin{verbatim}
static int btree_open(...)
{
  ...
  //sprintf(zBuf,"0x%
  sprintf(zBuf,"%
  ...
}
...
static int btree_pager_stats(...)
{
  ...
  //if(Tcl_GetInt(interp, argv[1], (int*)&pBt))
  //  return TCL_ERROR;
  pBt = strtol(argv[1], NULL, 16);
  if(!pBt) return TCL_ERROR;
  ...
}
\end{verbatim}
}

The \texttt{SEGFAULT} error continues to show up during the testing
process. Further gdb traces reveal that the \texttt{src/sqliteInt.c}
source code also assumes that pointer variable can be represented by
an \texttt{int} variable which does not hold true in 64 bit
systems. Therefore, code is updated as follows.  {\small
\begin{verbatim}
//# define INTPTR_TYPE int
# define INTPTR_TYPE long

/* Big enough to hold a pointer */
typedef INTPTR_TYPE ptr;
typedef unsigned INTPTR_TYPE uptr;
\end{verbatim}
}

All the tests runs successfully to completion after applying these
fixes.

\section{Critique of SQLite-2002}
\label{sec:critique}
A version of the SQLite database engine that was released 20 years ago
was compiled and tested on a modern hardware and software
platform. Several issues were encountered during this
exercise. Developing a fix for those issues provided valuable insight
into the factors that contribute to intelligibility and
maintainability of a program code and its documentation. This section
shows some of the key ideas in software engineering that emerges from
the software resurrection exercise.

\subsection{Change is the only constant in a software.}
A software tool operates in an ecosystem created by hardware
(e.g. CPU, memory, etc.), operating system and software
libraries. This ecosystem is continually changing in order to address
the requirements of the changing world. Therefore, change is the only
constant also in the life of a software. It is wiser to accept and
embrace the fact that changes to a software will be necessary as it
moves forward in time.

\begin{displayquote}
  ``Everything changes and nothing stands still.'' -- Heraclitus (a
  Greek philosopher)
\end{displayquote}

A class of updates to a software that will prevent normal operation of
other software tools or services that depends on the software is
called a \textit{breaking change}. While a breaking change is
undesirable, it is often essential. The Issue~\ref{compile:issue1} has
revealed that it is important to have flags or markers that caution
the users of such breaking changes at the point of usage. The GCC
compiler developers have wisely chosen to include a \texttt{varargs.h}
file in all GCC compiler distributions -- since 2004 -- which produces
an informative error message when the compiler attempts to use the
unsupported \texttt{varargs.h} header file.

{\small
\begin{verbatim}
$ cat /usr/lib/gcc/.../include/varargs.h
#ifndef _VARARGS_H
#define _VARARGS_H

#error "GCC no longer implements ."
#error "Revise your code to use ."

#endif
\end{verbatim}
}

Posting critical information at the point of usage is an important
construct for introducing a breaking change. In the case of compilers,
this involves showing an informative error message when a user tries
to access an unsupported feature. Further details about a breaking
change can also be disseminated through other forms of communication
like mailing list, software release document, etc. For example, the
GCC compiler release document contains a clear and concise notice
about this breaking change.
\begin{displayquote}
  ``GCC no longer ships \texttt{<varargs.h>}. Use \texttt{<stdarg.h>} instead.'' --
  \href{https://gcc.gnu.org/gcc-3.4/changes.html}{GCC 3.4 Release
    Series}
\end{displayquote}

The developers of the Tcl library could not provide information about
a breaking change at the point of usage. To find a fix for
Issue~\ref{test:issue1}, a maintainer has to explore the software
documentation. A more intelligible compiler error message suggesting a
corrective action would have been more useful. For example, a software
trying to access the \texttt{result} field of the \texttt{Tcl\_Interp}
data structure should be informed with an error message that this
feature is no longer available without defining the macro
\texttt{USE\_INTERP\_RESULT}. The GCC compiler permits, for example,
the \texttt{deprecated} attribute for a function to show a warning if
the an unsupported function gets used. However, such a feature is not
available for data member access and is the reason, most likely, why
such informative warnings were not generated when the \texttt{result}
field of the \texttt{Tcl\_Interp} data structure was accessed.

\subsection{To depend or not to depend, is a profound question that
  the wise can answer.}

The SQLite developers chose to rely on a non-standard feature
(i.e. vararg.h) provided by the compilers of their time (i.e. year
2002). The Issue~\ref{compile:issue1} revealed that dependence on
non-standard features makes the software vulnerable to changes in
ecosystem thereby increasing maintenance costs.

Issue~\ref{test:issue1} revealed another fact about software
dependencies; if software A depends on a software library B then it
implies that A has accepted that its fate is tied with the fate of
B. The SQLite software uses the TCL library to implement its test
suite. One can understand the benefit of this dependence; it allows
the SQLite developers to easily write tests in the Tcl scripting
language which is more concise, clear and easier to maintain. The cost
of such dependence is often overshadowed by the benefits. All
dependencies have a cost and understanding the cost is the first step
in taking a wise decision on whether to depend on a third party
library or to develop your own functionality. ``To depend or not to
depend is the question my dear developers'', would have asked
Shakespeare if he were reflecting on the pros and cons of software
dependencies. A wise developer will look at the benefits and costs of
a software dependency with certain degree of impartiality in order to
truly evaluate the impact of such dependencies.

\subsection{Unitary (or atomic) revisions to a version control system
  are more useful.}

Software developers often use a Version Control System (VCS) to keep a
historical record of changes (or, revisions) being made to a
software. Such historical record of revisions not only helps
understand the growth of a software and its structure but also allows
changes to be removed with surgical precision when the software
behaves in undesirable ways. It is easier to understand and reason
about a revision that introduces only one conceptual change (\eg a
feature, a bug fix, a new test case, \etc) in a software. Such
revisions can be said to be unitary (or atomic) as they reflect a unit
of change in the software.

The value of unitary revisions were realised while fixing the
Issue~\ref{compile:issue2}. The \texttt{getline()} identifier name
conflict with the standard library was fixed as a
\href{https://sqlite.org/src/vpatch?from=93a2c961b17d2459&to=558969ee8697180c}{single
  revision} in the version control history of the SQLite. This
revision contained relevant keywords (\eg \texttt{getline}) in the
revision log message that made it easier to locate. Such a unitary
revision was not available for Issue~\ref{compile:issue1} whose fix
was more difficult to develop as the fix required manually selecting
code updates from one of the
\href{https://www.sqlite.org/src/info/590f963b6599e4e2}{future
  revisions}. Therefore, unitary (or atomic) revisions with a revision
log containing all the relevant keywords are useful. When deciding
about the keywords relevant for a revision, it helps to think about a
learner who is allowed to search the revision log history using only
one or two keywords.

\subsection{Global Identifier Names Should be both Unique and Intelligible}

The developers of sqlite-2002 came up with the \texttt{getline()}
method name well before the method was defined in the standard library
through the \texttt{<stdio.h>} header file. They fixed this name
conflict by renaming the method to \texttt{local\_getline()} which not
only avoided conflict with the \texttt{getline()} method in the
standard library but also avoided any future conflicts with other
software. This was a wise decision because the updated method name has
survived more than 20 years of change in compilers and standard
libraries.

In some programming languages, like the C programming language, the
identifier names (e.g. function names, variable names, etc.) are
stored in either global or local scope. Names stored in the global
scope (e.g. function names such as \texttt{getline()}) are accessible
to all parts of the program and therefore have greater chance of
conflicting with other parts of the software (e.g. standard
library). Identifiers in the local scope (e.g. variable names defined
within a function) are only accessible in that local scope and
therefore has smaller chance of conflict with other identifiers
defined in that scope. Here is an example, taken from the SQLite code,
showing identifiers in both global and local scope.

{\small
\begin{verbatim}
static char *getline(char *zPrompt, FILE *in){
  char *zLine;
  int nLine;
  ...
}
\end{verbatim}
}

In the above code snippet, the function name \texttt{getline()} is
stored in the global scope and therefore it can be invoked from any
part of the program including the local scope of any other
function. On the other hand, the variable \texttt{zPrompt},
\texttt{in}, \texttt{zLine}, \texttt{nLine} are stored in the local
scope of the \texttt{getline()} function and therefore these variables are
only accessible from within the \texttt{getline()} function.

\begin{figure}[ht]
  \centering
  \includegraphics[width=0.6\linewidth]{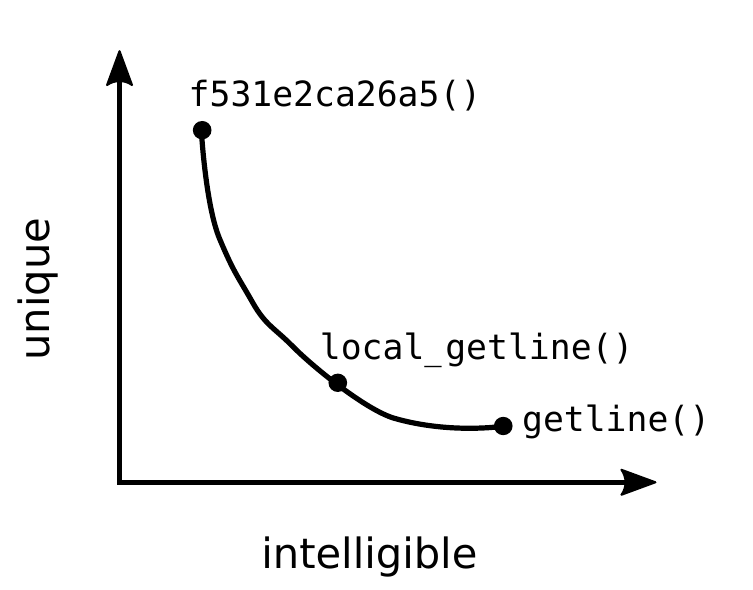}
  \caption{Identifier names (e.g. function name, variable name, etc.)
    in the global scope should be both unique and intelligible. Only
    unique or only intelligible identifier names are less useful.}
  \label{fig:intelligible-and-unique-global-identifiers}
\end{figure}

Programmers have to deal with the following two conflicting
requirements when crafting the identifier names that have to live in
the global scope.

\begin{itemize}
\item \textit{Uniqueness} : The identifier names must to be unique so
  that it does not conflict with identifiers that are currently
  defined and the identifiers that will get defined in the future in
  other parts of the software (e.g. standard library, dependent
  libraries, etc.). So the challenge for the programmer is to think of
  an identifier name that cannot be possibly thought by anyone else.
\item \textit{Intelligible} : The identifier names should be
  intelligible in order to convey the purpose of that identifier to
  the reader. The more informative an identifier name is, the more
  useful it is to the reader.
\end{itemize}

If one chooses a fairly unique name (e.g. f531e2ca26a5) then it is
unlikely that this identifier name will ever conflict with any other
identifiers. However, such identifier names will not be intelligible
to the readers as it does not convey any meaning. On the other hand,
one can chose a highly intelligible identifier name
(e.g. \texttt{getline()}) that fully conveys the purpose of the
method. However, such an intelligible name would also invite conflict
as it is highly likely that others will also want to use such an
intelligible name for a function that has a similar
purpose. Therefore, one has to chose an identifier name that lives in
the global scope by carefully balancing the uniqueness and
intelligibility requirements. The developers of SQLite chose such a
balance by adding the prefix \texttt{local} to the method name.

Many programming languages have introduced the concept of namespace to
compartmentalise global identifier names into buckets (or, namespace)
such that these identifiers can only be accessed using the name of the
bucket. For example, in the C++ programming language the
\texttt{getline()} method defined by the standard library would be
placed in the std namespace and be accessed using
\texttt{std::getline()} identifier name. The getline() function
defined by the SQLite database engine would be placed in the sqlite
namespace, for example, and would be accessed using
\texttt{sqlite::getline()} identifier name. One could argue that this
is similar to the practice of adding a prefix (or suffix) to an
identifier name in order to avoid conflicts. This is a valid
argument. However, a feature provided by the compiler makes the
concept of namespace more useful in reducing conflict and improving
intelligibility. Namespace have a scope and if an identifier is used
within a namespace, then one can remove namespace name prefix from the
identifier name. The default compiler behaviour is to assume that all
identifier names without the namespace prefix correspond to the
current namespace. This feature helps to improve intelligibility while
maintaining uniqueness. However, the reader is required to be aware of
the current namespace in order to correctly resolve an identifier
name.

\subsection{Tests build trust}
The SQLite database engine was ready to use after the successful
compilation of the sqlite-2002 as described in
Section~\ref{sec:compile}. One could create some tables, insert some
sample data, search for an entry and delete those entries to confirm
that the database engine is functioning properly even after 20 years
of its release. While these mock tests allow verification of the basic
functionality of the software, it is not sufficient to build trust in
the software. In other words, the ad hoc and mock tests cannot
convince someone to use this software for storing and managing
critical information like financial or medical data. One requires more
stronger guarantees which can only be provided by automated and self
contained tests that can be run by the users in their own
machines. The developers of the sqlite-2002 are best placed to build
such automated tests because they can reasonably be expected to
quantify the desired behaviour of the software. Such suite of software
tests are essential to build trust between the user and the
software. The sqlite-2002 software contains such a self contained
suite of tests and therefore allows its users to build such a trust
even after 20 years of its release.

\subsection{Tests define the expected behaviour of a software.}

Successful compilation of a 20 year old software (\ie sqlite-2002) on
a modern platform brought happiness and a sense of
achievement. However, these feelings were quickly overshadowed by
doubts as it became clear that the sqlite-2002 software was not
designed to run on future platforms; the developers of SQLite could
not be foreseen how the hardware and software would change in 20
years. Therefore, although the software compiled successfully in a
modern platform, it was not known if the software can function in the
way that its developers had expected during its released in the year
2002. Is it possible to concretely define the behaviour of a software
as it was desired in the year 2002?

The sqlite-2002 documentation states that this version of sqlite,
``implements a large subset of SQL92.'' and allows ``atomic commit and
rollback protect data integrity''. It is possible to prepare some SQL
query statements based on the SQL92 standard and prepare some SQL
tables to test atomic commits. However, will these be sufficient to
confirm that the software is truly behaving in the way it was designed
to operate before its release in the year 2002? The developers of
sqlite-2002 were best placed to know the desired and expected
behaviour of this version of SQLite. One could write hundreds of pages
of documents to describe the expected behaviour of a
software. However, the only quick and easy way to verify the
functionality of a software is to define the expected behaviour of the
software in the form of automated and self contained tests. These
tests not only verify functionality of a software on a new platform
but also acts as a concrete specification of the correct behaviour of
a software. Such tests are also useful for developing an understanding
of the input and output characteristics of a software.

\subsection{Who tests the Test?}
An automated and self contained test suite is a program code that has
been designed to test a software. The test suite invokes various
features of the software with a set of test inputs and compares it
against a set of corresponding test outputs. If the test suite has a
complex logic and involves substantial amount of program code then the
test suite merits a testing process for itself to ensure that the test
code does not have any flaws. This leads to a recursive testing
dependency in which a tests suites is developed to test another test
suite. Such a never ending scenario can only be avoided by a test
suite that is so simple that it does not demand a test for itself. A
simple test suite has minimal code and has minimal chances of
failure. Only such a simple test suite is capable of assuring that a
failed test case points to a failure in the software being tested and
does not correspond to a flaw in the test code itself.

\begin{figure}[ht]
  \centering
  \includegraphics[width=\linewidth]{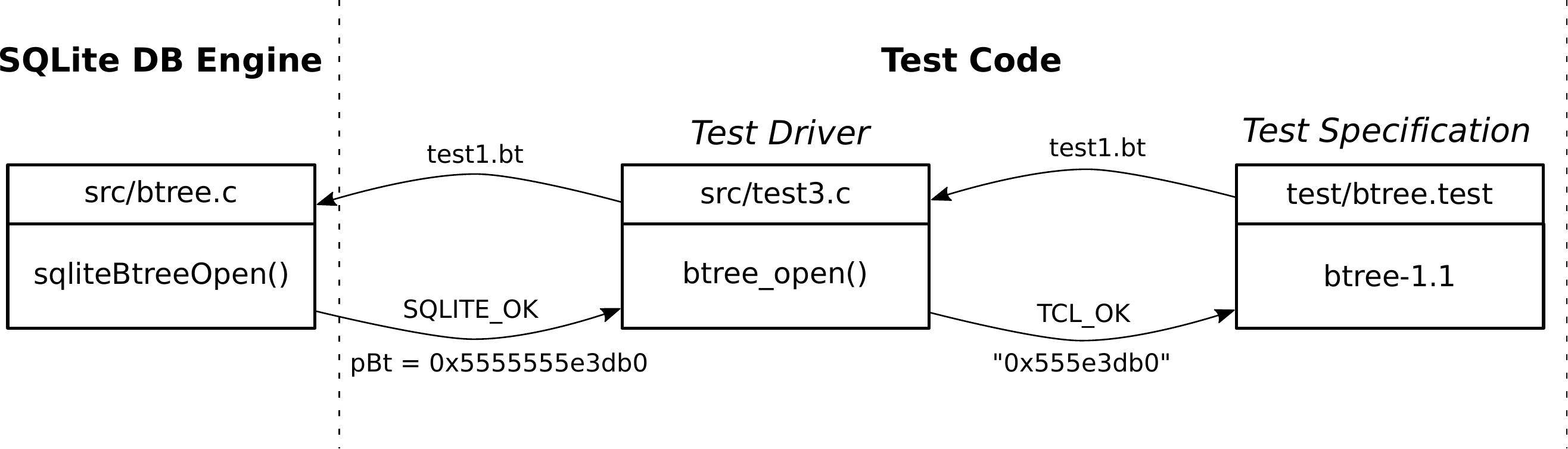}
  \caption{Illustration of control and data flow for test executions
    in sqlite-2002. The test specifications are defined in the Tcl scripting
    language. The sqlite core database engine uses the C programming
    language. A driver layer exposes the functionality of the sqlite core
    database engine library  which exposes different
    functionality of the core sqlite database engine as Tcl
    commands. }
  \label{fig:test-distance}
\end{figure}

The test cases (\eg \texttt{btree-1.1}) in sqlite-2002 are defined
using the Tcl programming language (\eg \texttt{test/btree.test}). The
core SQLite database engine is defined using the C programming
language (\eg \texttt{src/btree.c}). A test driver layer contains a
set of functions (\eg \texttt{btree\_open} defined in
\texttt{src/test\{1,2,3\}.c}) that allows the Tcl based test
specifications to access functionality of the core SQLite database
engine. \figurename~\ref{fig:test-distance} illustrates the control
and data flow for the test case \texttt{btree-1.1} which verifies the
ability of the sqlite engine to create a new database and represent it
using a btree data structure stored in a disk file.

The sqlite-2002 tests failed to run on a modern 64 bit platform
because the test driver layer assumed that memory addresses were 32
bits long. Such an assumption broke most of the test code because the
driver layer wrongly translated 64 bit pointer addresses (\eg
\texttt{0x5555555e3db0}) to 32 bit pointer addresses (\eg
\texttt{0x555e3db0}) by dropping the higher 32 bits portion of the
address. The test code did not operate properly and therefore failed
to properly test the software. The test control and data flow shown in
\figurename~\ref{fig:test-distance} must have its merits and therefore
was chosen by the sqlite-2002 developers. However, this failure forces
one to rethink about introducing any complexity in test suites and aim
for tests that are based on the most fundamental and stable features
of a compiler.

\begin{figure}[ht]
  \centering
  \includegraphics[width=\linewidth]{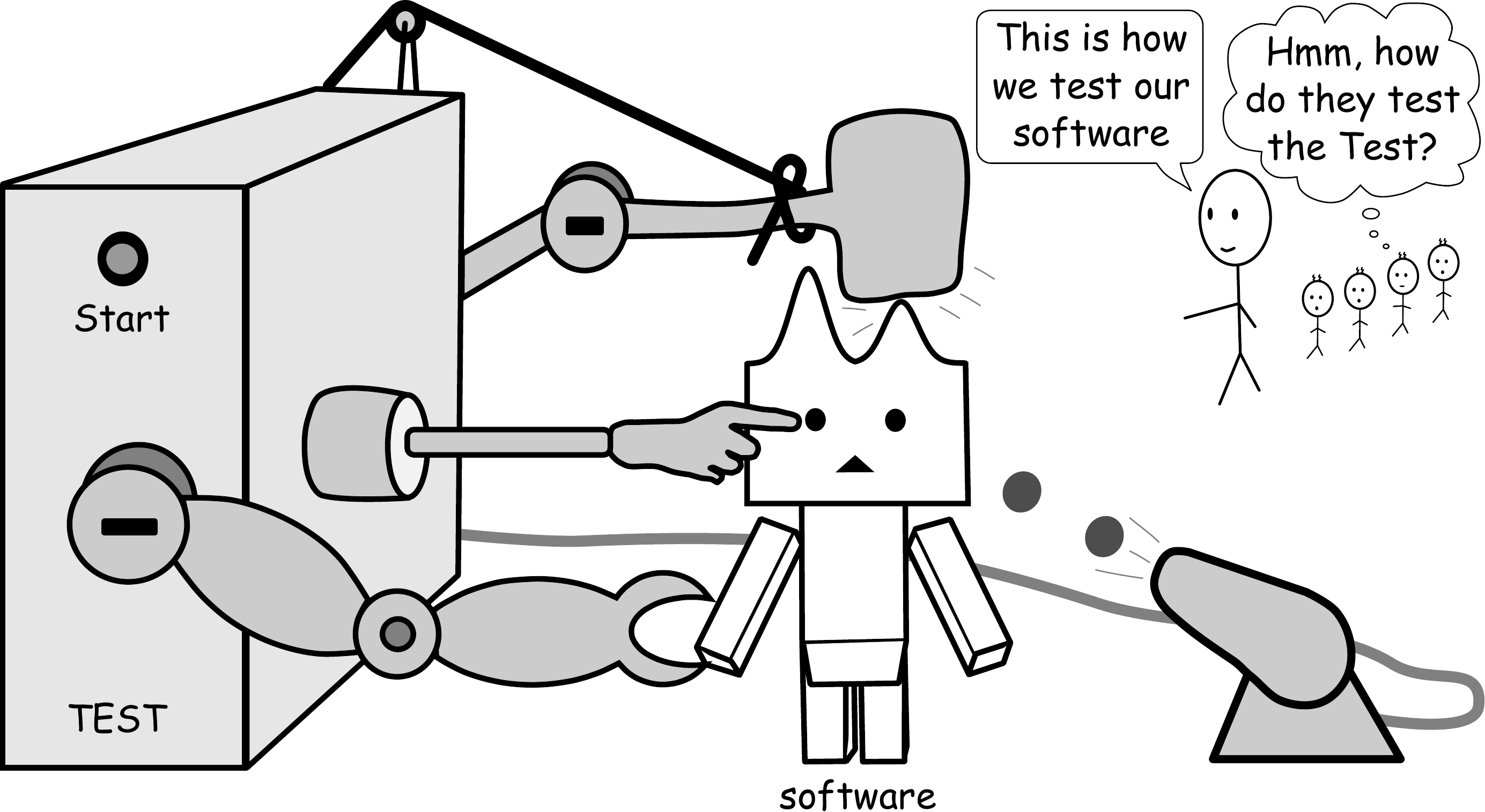}
  \caption{If the program code for testing a software is complex, then
    it demands a test for itself. A test suite is useful only when it
    is simple to operate and therefore does not merit a test
    for itself.}
  \label{fig:who-tests-the-Test-dilemma}
\end{figure}

\subsection{SQLite-2002 is an intelligible and maintainable software}
The SQLite database engine that was released 20 years ago can be
compiled in a modern hardware and software platform. It also delivers
all the features included in the original software release as
evidenced by successful execution of the regression tests. The program
code and regression tests required corrective software maintenance in
order to address the changes in the hardware and software
libraries. These maintenance activities could be performed by a modern
day developer because the program code is well documented. A
\texttt{README} text file provides an inviting introduction to the
software. Every function definition in the source code includes a
clear and concise description of its purpose. The user defined data
types are intelligible because of their identifier names as well as
the comments accompanying them. The software's architecture and
purpose of its various components are also well documented in a set of
self contained and static and offline HTML files that are generated by
the autoconf's \texttt{make doc} command. These documents and
comments, written 20 years ago, communicates to a modern day developer
with remarkable effectiveness and clarity. The sqlite-2002 software is
therefore an intelligible and maintainable software that is equally
valuable for learning some of the software engineering principles.

\section{Related Work}
\label{sec:related_work}

The software resurrection exercise provides an opportunity to engage
with historical program code and documentation. Such engagements with
things that are remote in time or space has been recognised by many as
a valuable learning experience. Harry R. Lewis~\cite{lewis2021ideas}
has compiled a book containing research papers from Computer Science
that have shaped the modern age. The goal of this book is to relieve
modern day readers of the ``misimpression that the established
conventions of the [Computer Science] field were handed down to
contemporary culture in finished form''~\cite{lewis2021ideas}.

Reading program code and documentation is an integral part of the
software resurrection exercise. The compilation and testing stages
requires a learner to read the program code and documentation in order
to understand the issues that prevents the software from operating in
a modern hardware and software platform. Such active engagement with
program code written by others has been recommended by many as a
method to improve one's ability to write maintainable and intelligible
software. For example, \cite{ryan2022reading} have designed a course
for a professional Master program in Software Engineering that teaches
students ``how to read the code of an existing, large-scale system to
become an effective contributing member of its community''.

The software resurrection exercise also facilitates the discovery of
programming pearls (or wisdom) by allowing learners to engage with
historical program code and documentation. Such active engagements
naturally evokes, in one's mind, the words of programming wisdom that
are often shared by experienced programmers. By virtue of their long
programming careers and their engagement with program code and
documents from a wide variety of application, experienced programmers
are able to come up with the programming wisdom that helps learners
understand the practice of computer programming. Jon Bentley collected
a set of such programming wisdom in a book titled ``Programming
Pearls'' and remarked that ``these programming pearls have grown from
real problems that have irritated real
programmers''~\cite{bentley2016programming}. Kernigham and Plauger
advised programmers to ``Write clearly -- don't be too
clever''~\cite{kernighan1974elements} possibly because they had gone
through the challenges of understanding a cleverly optimised code and
endured the cost of maintaining such unintelligible code. Donald
E. Knuth reflected similar sentiments by warning programmers that
``premature optimisation is the root of all evil (or at least most of
it) in
programming''~\cite{knuth1974computer-programming-as-an-art}. The
software resurrection exercise provides experiences and context that
allows a learner to truly understand the meaning of these programming
pearls.

\section{Discussion}
\label{sec:discussion}
This paper has introduced the concept of software resurrection as a
learning exercise. Historical release of the SQLite database engine,
released 20 years ago, was compiled and tested in a modern hardware
and software platform as a part of the software resurrection
exercise. Some of the issues (\eg issues~\ref{compile:issue1} and
\ref{compile:issue2}) encountered during this exercise were easy to
fix particularly because it was possible to borrow code from future
version control revisions of the SQLite code repository. These issues
requires few hours of exploration and learning. The
Issue~\ref{test:issue2} has been presented in this paper in a
condensed form but required many hours (at least 3 weekends) to fix
and was indeed a frustrating experience as the explorations often led
to a dead end. The software resurrection exercise does require a
motivated learner who has the perseverance to systematically
investigate an issue and logically reason about possible
solutions. The frustrating experiences contribute to learnings that
will potentially stay for a life long. Well written code and
documentation provides the joy that is similar to the emotions evoked
by a good novel, poem or an essay. The true value of self contained
and automated tests is realised when a 20 year old code says ``All
tests passed'' on a modern hardware and software platform that could
not have been foreseen by the past. The software resurrection exercise
has already helped at least one programmer but has the potential to
enlighten many more programmers.

\section{Conclusion}
\label{sec:conclusion}
The software resurrection exercise requires a learner to compile, test
and critique a historical released of a well maintained and widely
adopted software on a modern hardware and software platform. The
learner is required to fix issues encountered during the compilation
and testing process. These activities simulates the software
maintenance process and allows the learner to engage with program code
and documentation written by others. An example of the software
resurrection exercise provides early indications that it is a valuable
tool for learning the art of writing intelligible and maintainable
program code and documents. Further examples and experience reports
from prospective learners are required to scientifically validate this
claim.

\bibliographystyle{IEEEtran}
\bibliography{IEEEabrv,ref}

\end{document}